\documentclass[aps,preprint,prd,epsfig,nofootinbib]{revtex4-1}
\pdfoutput=1
\usepackage{graphicx}
\usepackage{longtable}
\usepackage{float}
\usepackage{dcolumn}
\usepackage{bm}
\usepackage{appendix}
\usepackage{multirow}
\usepackage{color}
\usepackage[utf8]{inputenc}
\usepackage{footmisc}
\usepackage{soul}
\usepackage{txfonts}
\usepackage{xcolor}
\usepackage{graphicx}
\usepackage{bm}
\usepackage{ulem}

\usepackage{amsmath}

\usepackage[colorlinks=true,linkcolor=blue,citecolor=blue]{hyperref}
\usepackage{feynmp-auto}
\usepackage{fancyhdr}
\usepackage{lineno}

\usepackage{wasysym}

\usepackage{xcolor}
\usepackage{graphicx}
\usepackage{dcolumn}
\usepackage{bm}
\usepackage{amssymb}
\usepackage{blindtext}
\usepackage{hyperref}
\begin{document}
\title { Scalar contributions from  $331RHN$ minimal model to  oblique parameters }
\author{A. Doff$^{\ddag}$\footnote[1]{\href{mailto:agomes@utfpr.edu.br }{\footnotesize agomes@utfpr.edu.br }}}

\affiliation{\footnotesize $^{\ddag}$ Universidade Tecnologica Federal do Parana - UTFPR - DAFIS, R. Doutor Washington Subtil Chueire, 330 - Jardim Carvalho, 84017-220, Ponta Grossa, PR, Brazil}
\date{\today}
\begin{abstract}
Electroweak precision observables, encoded in the oblique parameters $S$, $T$, and $U$, impose stringent constraints on extensions of the Standard Model. In this work, we analyze the scalar-sector contributions to these parameters within the minimal 331RHN model. Building on previous results obtained in the 331RHN framework, we show that the oblique parameter $T$ provides the dominant constraint on the scalar mass spectrum. Our results indicate that current experimental bounds on $T$ lead to a nontrivial upper limit on the symmetry-breaking scale, $\omega \lesssim 10~\text{TeV}$. These findings highlight the sensitivity of electroweak precision data to the scalar sector of 3-3-1 models and their viability as extensions of the Standard Model.
\end{abstract}
\maketitle
\section{Introduction}

\par Extensions of the Standard Model based on the gauge symmetry 
$SU(3)_C \times SU(3)_L \times U(1)_Y$  have long attracted attention as a compelling framework for addressing some of the open questions in particle physics. Early formulations of this class of models can be traced back to the work of Ref.~\cite{Singer:1980sw}, where extended electroweak gauge structures were first explored. The modern realization of this class of  models was later established independently in Refs.~\cite{Pisano:1992bxx,Frampton:1992wt}, where it was shown that embedding the electroweak sector into $SU(3)_L \times U(1)_Y$ leads to nontrivial constraints on the fermion content.

\par A particularly remarkable feature  is that the cancellation of gauge anomalies occurs only when the number of fermion generations is a multiple of three, with the simplest consistent realization corresponding to exactly three families. This provides an appealing theoretical explanation for the observed replication of fermion generations~\cite{Liu:1993gy,Pisano:1996ht}. In addition, the structure of the extended gauge group naturally leads to the quantization of the electric charge~\cite{deSousaPires:1998jc,deSousaPires:1999ca}, which emerges as a consequence of the gauge symmetry and fermion representations, rather than being imposed as an external condition.

\par Among the various realizations of the 3-3-1 framework, one of the most extensively studied is the model with right-handed neutrinos (331RHN)~\cite{Foot:1994ym}. In this construction, right-handed neutrino states are naturally accommodated within the lepton triplets, leading to a richer fermionic content and an enlarged scalar sector. A comprehensive review of the theoretical foundations of the model, together with its phenomenological developments over the past three decades, can be found in Ref.~\cite{Escalona:2025jla}.

\par Motivated by the search for more economical realizations of the symmetry-breaking mechanism, several 3-3-1 scenarios with reduced scalar sectors have been proposed over the years. Within the context of the 3-3-1 model~\cite{Pisano:1992bxx}, we can highlight the minimal 3-3-1 model with three scalar triplets~\cite{331-3trip-2010}, the reduced 3-3-1 model~\cite{331m1-2011}, and the simple 3-3-1 model~\cite{331red-2014}. These constructions retain the essential features of the 3-3-1 gauge structure while exhibiting distinctive theoretical and phenomenological properties arising from their simplified scalar sectors.

\par Along the same line, a minimal realization of the 331RHN framework\cite{Foot:1994ym}, hereafter referred to as the minimal 331RHN model, was proposed in Ref.~\cite{Ponce:2002sg}. Its phenomenological implications were subsequently investigated in detail in Ref.~\cite{Dong:2006xf}, where the gauge interactions and particle spectrum were systematically analyzed.

\par From a phenomenological standpoint, a characteristic prediction of 3-3-1 models is the unavoidable presence of flavor-changing neutral currents (FCNCs) at tree level, mediated by both scalar and gauge bosons~\cite{Long:1999ij,Buras:2014yna,Oliveira:2022vjo,Escalona:2025rxu}. In addition, both sectors give rise to nontrivial contributions to electroweak precision observables, particularly to the oblique parameters. In a recent study~\cite{331STU:2026dc}, we analyzed the scalar-sector contributions to the oblique parameters \(S\), \(T\), and \(U\) within the 331RHN model. A related analysis was carried out in Ref.~\cite{STM331}, where the implications of the oblique parameters were investigated in the version of the model proposed in Ref.~\cite{Pisano:1992bxx}. Our results showed that the parameter \(T\) provides the most stringent constraints on the scalar mass spectrum and the associated symmetry-breaking scales, leading to the bounds \( v_{\chi^\prime} \lesssim O(14)\,\text{TeV}\) and \( f \lesssim O(10)\,\text{GeV}.\)

\par Building upon these results, we extend the previous analysis to the minimal 331RHN model in order to investigate whether the current experimental bounds on the oblique parameter  \(T\) can also constrain the symmetry-breaking scale \(\omega\). As we will show, the parameter  \(T\)  remains the dominant electroweak precision observable in this framework and leads to a nontrivial upper bound on \(\omega\).

\par This work is organized as follows. In Sec.~II, we outline the essential features of the model that form the basis of our analysis. In Sec.~III, we evaluate the scalar contributions to the oblique parameters and present the corresponding numerical results. Finally, in Sec.~IV, we summarize our main findings and provide concluding remarks.

\section{Main aspects of the model}

\subsection{Fermion content}
In this model, the leptonic sector is analogous to that of the 331RHN model~\cite{Foot:1994ym}, and consists of triplets and singlets under  $SU(3)_C \times SU(3)_L \times U(1)_N$,
\begin{eqnarray}
L_{aL}=
\left(
\begin{array}{c}
\nu_a \\
e_a \\
\nu^c_a
\end{array}
\right)_L
\sim (1,3,-1/3),
\qquad
e_{aR}\sim(1,1,-1),
\label{lc}
\end{eqnarray}
where $a=e,\mu,\tau$ labels the three Standard Model lepton generations.

The quark sector consists of three families and, in order to ensure the cancellation of chiral anomalies, one quark family must transform differently from the other two under the 3-3-1 gauge symmetry. In the the 331RHN minimal model realization, the quark fields transform as
\begin{eqnarray}
&&Q_{1L}=
\left(
\begin{array}{c}
u_1 \\
d_1 \\
U_1
\end{array}
\right)_L
\sim(3,3,1/3),
\qquad
u_{1R}\sim(3,1,2/3),
\nonumber\\
&&d_{1R}\sim(3,1,-1/3),
\qquad
U_{1R}\sim(3,1,2/3), \nonumber \\
&&Q_{i L}=
\left(
\begin{array}{c}
d_i \\
-u_i \\
D_i
\end{array}
\right)_L
\sim (3,\bar{3},0),
\qquad
\,\,\,\,\,\,u_{iR}\sim(3,1,2/3),
\nonumber\\
&&d_{iR}\sim(3,1,-1/3),
\qquad
D_{iR}\sim(3,1,-1/3),
\label{quarks}
\end{eqnarray}
where the index $i=2,3$ runs over the two second families transforming as anti-triplets. The minus sign in the anti-triplet $Q_{iL}$ is introduced as a convention to reproduce the standard structure of charged current interactions with the gauge bosons. In the equations above, $U_1$ and $D_i$ denote new heavy quarks carrying the usual electric charges $(+2/3,-1/3)$.

After the spontaneous symmetry breaking ${G}_{331}\to{ G}_{\rm SM}$, the fermionic triplets decompose into the Standard Model doublets plus additional singlet states. As a result, no new chiral electroweak doublets are introduced beyond those already accounted for in the gauge structure, and therefore the fermionic sector does not generate additional independent contributions to the oblique parameters beyond the effects already encoded in the extended gauge and scalar sectors.
\subsection{Scalar sector}

The scalar sector of  the 331RHN model with minimum scalar sector differs significantly from that usual 331RHN model. In particular, the spontaneous symmetry breaking of  $SU(3)_C \times SU(3)_L \times U(1)_N$ is achieved in two stages through a minimal scalar content consisting of only two scalar triplets.

At the first stage, the gauge symmetry is broken down to the Standard Model group through the scalar triplet $\chi = (\chi^0_1, \chi^{-}_2, \chi^{0}_3)^T \sim (1,3,-1/3)$,  whose neutral components $\chi_1^0$ and $\chi_3^0$ develop vacuum expectation values (VEVs). At the second stage, the electroweak symmetry is broken down to the electromagnetic group $U(1)_Q$ through the scalar triplet $\phi = (\phi^+_1, \phi^{0}_2, \phi^{+}_3)^T \sim (1,3,2/3)$,   with the neutral component $\phi_2^0$ acquiring a VEV.

\par At this point, it is important to emphasize that in the triplet $\chi$, 
 its neutral components $\chi^0_1$ and $\chi^0_3$ acquire vacuum expectation values of order  $u \sim O(\mathrm{GeV})$ and  $\omega \sim O(\mathrm{TeV})$, respectively, establishing a hierarchical structure in the symmetry-breaking pattern.
\par In this framework, the large VEV $\omega$ generates masses for the exotic quarks $U_1$ and $D_i$, while the smaller VEV $u$ contributes to the masses of the ordinary quarks $u_1$ and $d_i$. On the other hand, the electroweak VEV $v$ developed by $\phi_2^0$, is responsible for generating the masses of the remaining ordinary quarks, namely $u_i$ and $d_1$, as well as all ordinary leptons, as discussed in~\cite{Dong:2006xf}.

\par With this scalar content, the most general renormalizable Higgs potential takes the simple form
\begin{eqnarray}
V(\phi,\chi)&=&
\mu_1^2(\chi^\dagger\chi)
+\mu_2^2(\phi^\dagger\phi)
+\lambda_1(\chi^\dagger\chi)^2
+\lambda_2(\phi^\dagger\phi)^2
\nonumber\\
&&
+\lambda_3(\chi^\dagger\chi)(\phi^\dagger\phi)
+\lambda_4(\chi^\dagger\phi)(\phi^\dagger\chi).
\label{potential}
\end{eqnarray}
An important feature of this version of the 3-3-1 model is the absence of trilinear scalar interactions, which makes the Higgs potential considerably simpler than in the 331RHN model.

After spontaneous symmetry breaking, the neutral scalar fields are shifted around their VEVs as
\begin{equation}
\phi^0_2=
\frac{1}{\sqrt{2}}
\left(
v+R_{\phi^0_2}+iI_{\phi^0_2}
\right),
\label{vev1}
\end{equation}
\begin{equation}
\chi^0_1=
\frac{1}{\sqrt{2}}
\left(
u+R_{\chi^0_1}+iI_{\chi^0_1}
\right),
\label{vev2}
\end{equation}
\begin{equation}
\chi^0_3=
\frac{1}{\sqrt{2}}
\left(
\omega+R_{\chi^0_3}+iI_{\chi^0_3}
\right).
\label{vev3}
\end{equation}
Minimization of the scalar potential then leads to the stationarity conditions
\begin{eqnarray}
2\lambda_1(u^2+\omega^2)
+\lambda_3v^2
+\mu_1^2
&=&0,
\nonumber \\
\lambda_3(u^2+\omega^2)
+2\lambda_2v^2
+\mu_2^2
&=&0.
\label{mincond}
\end{eqnarray}
The mass matrix for the CP-even neutral scalars in the basis $(R_{\phi^0_2} ,R_{\chi^0_1}, R_{\chi^0_3})$  corresponds to 
\begin{equation}
M^2_R
=
2\begin{pmatrix}
2 v^2 \lambda _2 & u v \lambda _3 & v \omega  \lambda _3 \\
 u v \lambda _3 & 2 u^2 \lambda _1 & 2 u \omega  \lambda _1 \\
 v \omega  \lambda _3 & 2 u \omega  \lambda _1 & 2 \omega ^2 \lambda _1 
\end{pmatrix},
\end{equation}


\par where this matrix   has zero determinant.

\par  From this , we can identify an exact massless eigenstate, which corresponds to the  Goldstone boson $G_1$~\cite{Ponce:2002sg},  that is given by 
\begin{equation}
    G_1= -\dfrac{\omega}{\sqrt{\omega^2 + u^2}}R_{\chi^0_1} +\dfrac{u}{\sqrt{\omega^2 + u^2}}R_{\chi^0_3},  
\end{equation}
\noindent while  the remaining two non-zero eigenvalues, corresponding to the physical CP-even Higgs states $h$ and $H_m$. 
 \par After diagonalization,  this  CP-even eigenstates   can be  identified by 
\begin{align}
  &h=  -\dfrac{\omega\kappa_1}{\sqrt{\omega^2\kappa^2_1 + v^2}}R_{\phi^0_2} + \dfrac{u}{\sqrt{\omega^2 + u^2}}\dfrac{v}{\sqrt{\omega^2\kappa^2_1 + v^2}}R_{\chi^0_1} + \dfrac{v}{\sqrt{\omega^2\kappa^2_1 + v^2}}\dfrac{\omega}{\sqrt{\omega^2 + u^2}}R_{\chi^0_3} \label{eqh}\\ 
  &H_m=  \dfrac{v}{\sqrt{\omega^2\kappa^2_1 + v^2}}R_{\phi^0_2} + \dfrac{u}{\sqrt{\omega^2 + u^2}}\dfrac{\omega\kappa_1}{\sqrt{\omega^2\kappa^2_1 + v^2}}R_{\chi^0_1} +\dfrac{\omega}{\sqrt{\omega^2 + u^2}}\dfrac{\omega\kappa_1}{\sqrt{\omega^2\kappa^2_1 + v^2}}R_{\chi^0_3}\label{eqH}, 
\end{align}
\noindent where  $\kappa_1 =2\lambda_1/\lambda_3$. The expressions for the masses of these scalars correspond to
\begin{eqnarray}
&& m_h^2 =  2 \left(\lambda _1 \left(u^2+\omega ^2\right)+\lambda _2 v^2-\sqrt{\left(\lambda _1 \left(u^2+\omega ^2\right)-\lambda _2 v^2\right){}^2+\lambda _3^2 v^2 \left(u^2+\omega ^2\right)}\right) 
\nonumber \\
&& m_{H_m}^2 = 2 \left(\lambda _1 \left(u^2+\omega ^2\right)+\lambda _2 v^2+\sqrt{\left(\lambda _1 \left(u^2+\omega ^2\right)-\lambda _2 v^2\right){}^2+\lambda _3^2 v^2 \left(u^2+\omega ^2\right)}\right) .
\label{mhH}
\end{eqnarray}
\noindent 
\par We now turn to the CP-odd scalar sector. By expanding the scalar fields around the vacuum expectation values (VEVs) defined in Eqs.~(\ref{vev1})--(\ref{vev3}), and substituting these expressions into the scalar potential in Eq.~(\ref{potential}), together with the minimization conditions of Eq.~(\ref{mincond}), one finds that no physical CP-odd pseudo-scalars remain in the spectrum. Instead, the fields
\(
I_{\phi^0_2},
I_{\chi^0_1},
\text{and }
I_{\chi^0_3}
\)
are identified as Goldstone bosons $G_2\equiv G, G_3$ and $G_4$,  as discussed in Ref.~\cite{Ponce:2002sg}.
\par Finally, with regard to  the charged scalars, assuming the basis defined by $\left(
\chi_2^+,\, \phi_1^+,\, \phi_3^+ \right)$,  the mass matrix is given by:

\begin{equation}
M_+^2
=
2\lambda_4
\begin{pmatrix}
v^2 & vu & v\omega \\
uv & u^2 & u\omega \\
v\omega & u\omega & \omega^2
\end{pmatrix}.
\label{M+}
\end{equation}
\par The  matrix listed above can be diagonalized by assuming the rotation parameterized by
\begin{equation}
\begin{pmatrix}
H^+_{m} \\
G_5^+  \\
G_6^+
\end{pmatrix}
=
\begin{pmatrix}
\cos\phi &
\sin\phi\,\sin\theta &
\sin\phi\,\cos\theta
\\[6pt]
0 &
\cos\theta &
-\sin\theta
\\[6pt]
-\sin\phi &
\cos\phi\,\sin\theta &
\cos\phi\,\cos\theta
\end{pmatrix}
\begin{pmatrix}
\chi_2^+ \\
\phi_1^+ \\
\phi_3^+
\end{pmatrix}
\label{at}
\end{equation}
where  we defined
\begin{eqnarray}
&& \sin\theta
=
\frac{u}{\sqrt{\omega^2+u^2}},\,\,\,\,
\qquad
\cos\theta
=
\frac{\omega}{\sqrt{\omega^2+u^2}},\nonumber \\
&& 
\sin\phi
=
\frac{\sqrt{\omega^2+u^2}}
{\sqrt{\omega^2+u^2+v^2}},\!\!\!\!\!\!\!
\qquad
\cos\phi
=
\frac{v}
{\sqrt{\omega^2+u^2+v^2}}.
\end{eqnarray}

\par The mass matrix defined in Eq.~(\ref{M+}) possesses two zero eigenvalues, corresponding to the two Goldstone bosons \(G_5^+\) and \(G_6^+\). From the rotation matrix defined in Eq.~(\ref{at}),  these states can be identified as
\begin{eqnarray}
 && G_5^+ = \cos\theta\, \phi_1^+ - \sin\theta\, \phi_3^+  \nonumber \\
 && G_6^+ = -\sin\phi\, \chi_2^+ +  \cos\phi\,\sin\theta \phi_1^+ +  
 \cos\phi\,\cos\theta \phi_3^+ ,
\end{eqnarray}
\noindent where \(G_5^+ \equiv G^+\) is  the Goldstone boson associated with the Standard Model Higgs doublet,  absorbed by \(W^+\), while \(G_6^+ \equiv G^{\prime +}\) is similarly absorbed by \(Y^+\). 

The remaining heavy charged scalar \(H^+_{m}\) can be identified as
\begin{eqnarray}
 && H^+_{m} = \frac{v}{\sqrt{\omega^2+u^2+v^2}}\chi^+_2 +  \frac{u}{\sqrt{\omega^2+u^2+v^2}} \phi^+_1 +  \frac{\omega}{\sqrt{\omega^2+u^2+v^2}}\phi^+_3,
 \label{massH+}
\end{eqnarray}
\noindent  whose mass corresponds to
\begin{equation}
 m^2_{H^+_{m}} = \lambda_4(\omega^2+u^2+v^2).  
 \label{m+1}
\end{equation}

In summary, the scalar spectrum of the 331RHN model with a minimal scalar sector contains only four physical scalar states: the CP-even neutral scalar \(h\), identified with the Standard Model-like Higgs boson; the heavy CP-even neutral scalar \(H_m\), and the  heavy charged scalars $H^{\pm}_{m}$. 

\subsection{Gauge Sector}

\par The gauge sector of the 331RHN model with a minimal scalar content closely resembles that of the conventional 331RHN framework. It consists of eight gluon fields associated with the gauge symmetry \(SU(3)_C\), eight electroweak gauge bosons corresponding to \(SU(3)_L\), namely \(W_1^\mu,\dots,W_8^\mu\), and an additional neutral gauge boson \(W_N^\mu\) linked to the \(U(1)_N\) factor.
\par Following spontaneous symmetry breaking, the gauge fields of the \(SU(3)_L \times U(1)_N\) sector undergo mixing, giving rise not only to the four electroweak gauge bosons of the Standard Model but also to a set of additional heavy vector states, namely \((Y^\pm, X^0, X^{0*}, Z')\). A comprehensive discussion of the vector boson spectrum in this framework can be found in Ref.\cite{Dong:2006xf}. In particular, it is worth emphasizing that the effects of these new gauge bosons on the oblique parameters \(S\), \(T\), and \(U\) were investigated in Ref.~\cite{Long:1999bny}.

\par After the symmetry-breaking pattern
\(
{ G}_{331} \to { G}_{SM},
\)
the bileptons organize themselves into an \(SU(2)_L\) doublet,
\(
(Y^+,X^0)^T.
\)
The main results of Ref.~\cite{Long:1999bny} indicate that the bilepton-induced contributions to the oblique parameters \(S\) and \(T\) become progressively suppressed as the bilepton masses increase, reflecting a characteristic decoupling behavior. In contrast, the contribution associated with the \(Z'\) boson becomes phenomenologically relevant for masses around \(10\) TeV, primarily due to \(Z\)-\(Z'\) mixing effects. Consequently, for new gauge bosons with masses at the TeV scale, the corrections to the oblique parameters remain negligible and will therefore be disregarded in the present analysis.


\section{Scalar contributions  of the 331RHN minimal model to the S, T, U parameters}

\subsection{Scalar doublets in the Higgs basis}

\par After the spontaneous breaking of the gauge symmetry
${G}_{331} \to {G}_{SM}$, the scalar triplets
$\chi$ and $\phi$ decompose under ${G}_{SM}$ into the following irreducible representations:
\begin{eqnarray}
\Phi_1 = \left (
\begin{array}{c}
\chi^{0}_1\\
\chi^{-}_2
\end{array}
\right ),\,\Phi_2 = \left (
\begin{array}{c}
\phi^+_1 \\
\phi^{0}_2
\end{array}
\right ),\,\, \chi^0_3 ,\, \phi^{+}_3,
\label{scalarcont}
\end{eqnarray}
where $\Phi_1$ and $\Phi_2$ transform as electroweak doublets, whereas $\chi_3^0$ and $\phi_3^+$ are singlets under ${G}_{SM}$. Following electroweak symmetry breaking, the charged singlet $\phi_3^+$ mixes with the charged components of the doublets, giving rise to the physical charged scalar spectrum, as described by Eq.~(\ref{massH+}). An analogous mechanism takes place in the neutral scalar sector, where the singlet field $\chi_3^0$ mixes with the neutral components of the doublets, leading to the physical neutral scalar states defined in Eqs.~(\ref{eqh}) and (\ref{eqH}).

\par Since physical observables must be independent of the particular choice of scalar basis, it is often advantageous to perform a basis transformation in field space. In particular, one may rotate to the Higgs basis~\cite{baseH}\cite{HiggsBase1}\cite{HiggsBase2}, in which the electroweak vacuum structure is entirely aligned along a single scalar doublet. In this basis, only $\Phi_2$ carries the full electroweak vacuum expectation value (VEV), while the orthogonal scalar combinations possess vanishing VEVs. This choice is especially useful because it cleanly separates the symmetry-breaking degrees of freedom from the additional physical scalar excitations, considerably simplifying both the interpretation of the scalar spectrum and the structure of its interactions.

\par Accordingly, in this basis the scalar doublets introduced in Eq.~(\ref{scalarcont}) are redefined as follows
\begin{equation}
\Phi_{2} =
\begin{pmatrix}
G^+ \\
\dfrac{1}{\sqrt{2}}(v + h + iG)
\end{pmatrix},
\qquad
\Phi_1 =
\begin{pmatrix}
H^+_{m} \\
\dfrac{1}{\sqrt{2}}H_m
\end{pmatrix}
\label{2tripl}
\end{equation}

\par The use of this basis is particularly well motivated when the objective is to systematically derive the contributions of scalar fields to gauge and matter currents. In this formulation, the couplings of the scalar sector acquire a particularly transparent structure: the Goldstone modes and the longitudinal components of the electroweak gauge bosons are entirely associated with the doublet $\Phi_2$, reflecting its exclusive role in electroweak symmetry breaking. Meanwhile, the additional scalar degrees of freedom encoded in $\Phi_1$  contribute only through interactions involving physical scalar states.

\par This separation proves especially advantageous in the analysis of scalar-induced corrections to gauge interactions, since it disentangles Standard Model-like contributions from  new-physics effects. In particular, it allows one to identify  how the extra scalar states contribute to charged and neutral currents, loop-induced processes, and electroweak precision observables.

\subsection{Scalar contributions from  $331RHN$ minimal model to  oblique parameters}

\par As mentioned in the introductory section, in Ref.~\cite{331STU:2026dc}
, we investigated the scalar contributions of the 331RHN model to the oblique parameters \(S\), \(T\), and \(U\). In that framework, the effective scalar spectrum responsible for additional contributions to the electroweak precision observables is organized into two distinct sectors. The first consists of the scalar set \((h^+_1, H, A)\), whose squared masses scale as \(\propto fv_{\chi^\prime}\) and may naturally lie close to the electroweak scale, depending on the mass scale set by the trilinear parameter \(f\). The second sector is formed by \((h^+_2, H^{\prime\prime})\), whose squared masses scale as \(\propto v_{\chi^\prime}^2\), placing these states naturally at the TeV scale. This hierarchical structure plays a central role in determining the pattern of scalar contributions to the oblique parameters.

\par As discussed in the previous subsection, the minimal 331RHN model exhibits a significantly simpler scalar structure. In this case, the effective scalar spectrum contributing to the oblique parameters consists only of the states \((H^+_{m}, H_m)\), whose squared masses are proportional to \(\omega^2\). Consequently, the electroweak corrections induced by the scalar sector are controlled primarily by the mass splitting between these states. In the degenerate limit,
\(
m_{H_m}=m_{H^+_{m}},
\)
the scalar contributions to the oblique parameters vanish identically, reflecting the characteristic suppression of precision-electroweak effects in the presence of degenerate scalar multiplets. 

\par Nevertheless, as demonstrated in ~\cite{331STU:2026dc}
, the oblique parameter \(T\) emerges as the most restrictive electroweak observable for the 331RHN model, allowing meaningful constraints to be placed on the mass scales defined by \(f\) and \(v_{\chi^\prime}\). In particular, we found that
\(
v_{\chi^\prime} \lesssim O(14)\,\text{TeV}
\)
and
\(
f \lesssim O(10)\,\text{GeV}.
\)
These results indicate that the \(T\) parameter provides a sensitive probe of the scalar mass structure, especially through effects associated with mass splittings and custodial-symmetry violation.

\par Motivated by this observation, and in order to establish whether the current experimental bounds on \(T\) may also impose a nontrivial constraint on the scale \(\omega\), in this subsection we evaluate the contributions of $\Phi_1$  to the weak isospin currents. This analysis is particularly relevant for determining whether deviations from the fully degenerate limit can induce observable corrections to electroweak precision parameters and, consequently, provide indirect bounds on the scalar sector of the minimal 331RHN framework.

\par The determination of the scalar-sector contributions to the vacuum polarizations in the minimal 331RHN model follows the same formal construction developed in Section III-B of  ~\cite{331STU:2026dc}, with an important simplification arising from the scalar spectrum of the present framework. In particular, due to the absence of an axion-like pseudoscalar state, the only nonvanishing scalar contributions relevant to the oblique analysis arise in the charged-current sector, namely through the polarization function $\Pi_{WW}$. As a consequence, the structure of the vacuum-polarization corrections is considerably simplified compared to  331RHN case. For a more detailed derivation of the formalism, we refer the reader to Section III-B of this work; below, we present the corresponding contribution of the doublets $\Phi_1$  to the polarization function $\Pi_{WW}$.

\begin{align}  
\Pi^{11}_{\mu\nu}(q)  & =  i\int d^4x\, e^{iq\cdot x}\Big(\langle J^1_\mu(x)J^1_\nu(0)\rangle_{ H_m H^+_{m}} \Big) = \left(q^2g_{\mu\nu} -  q_\mu q_\nu \right)\Pi_{11}(q^2),
\label{pola}
\end{align}

\noindent where, in the notation adopted here, $J_\mu^1$ denotes the weak isospin current associated with the $SU(2)_L$ generator $\sigma^1$, while the subscripts identify the scalar states circulating in the loop and specify the corresponding radiative corrections produced by $J_\mu^1$ that is given by
\begin{figure}[t]
\vspace*{-0.45cm}
    \centering
\hspace*{-0.2cm}\includegraphics[scale=0.7]{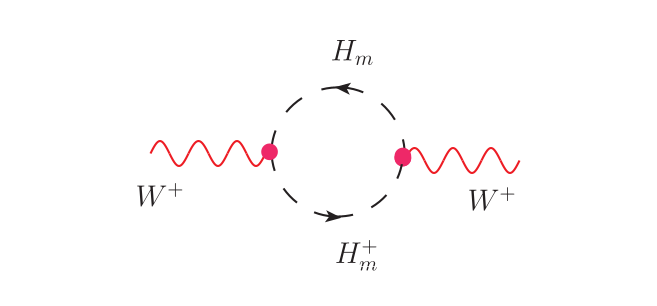}
    \caption{Feynman diagram  at  one-loop contributing  for  the vacuum polarization function defined in Eq.(\ref{pola}). }
    \label{fig1}
\end{figure}

\begin{align}
J_\mu^{1} & =  \frac{i}{2\sqrt{2}}\Big( H^+_{m}\hat{\partial}_\mu H_m
-H_m\hat{\partial}_\mu H^+_{m} \,\Big) + H.c. 
\label{331currA}
\end{align}
\noindent where  we define 
\begin{align}
\hat{\partial}_\mu H^+_{m}
&=
\partial_\mu H^+_{m}
- i g
\left(
\frac{1}{2} W_\mu^3H^+_{m}
+ W_\mu^+  H_m
\right)\label{partplus}\\
\hat{\partial}_\mu H_m
&=
\partial_\mu H_m
- i g
\left(
W_\mu^- H^+_{m}
- \frac{1}{2} W_\mu^3 H_m
\right)\label{partH} ,
\end{align} 
\noindent and the  diagrammatic realization is represented by the Feynman diagram showed in Fig.~1.

\par On the left-hand side of Eq.~(\ref{pola}), the superscript $(11)$ labels the scalar contribution to the gauge-boson polarization tensor in the custodial $SU(2)$ basis. These current-current correlators are directly related, through the standard current-algebra construction, to the gauge-boson self-energies entering the electroweak vacuum polarizations. More specifically, the transverse component of $\Pi^{11}_{\mu\nu}(q)$ determines the charged gauge-boson vacuum polarization $\Pi_{WW}(q^2)$, while the analogous quantity $\Pi^{33}_{\mu\nu}(q)$ is related to the neutral-current polarization $\Pi_{ZZ}(q^2)$. These quantities are ultimately evaluated in the limit of vanishing momentum transfer, $q^2 \to 0$, which is the relevant regime for the extraction of the oblique parameters.

\par The impact of additional scalar states on electroweak precision observables can be conveniently parametrized in terms of the oblique parameters $S$, $T$, and $U$, which are defined through the transverse components of the gauge-boson vacuum polarization amplitudes. In the $SU(2)_L \times U(1)_Y$ basis, these quantities are naturally expressed in terms of the correlators $\Pi_{11}(q^2)$ and $\Pi_{33}(q^2)$. In this framework, the oblique parameters take the form~\cite{PeskinTakeuchi1990,PeskinTakeuchi1992,AltarelliBarbieri1991} 
\begin{align}
\alpha S &= 4 s_W^2 c_W^2 
\left[
\Pi'_{33}(0) - \Pi'_{3Q}(0)
\right], \\
\alpha T &= 
\frac{\Pi_{11}(0) - \Pi_{33}(0)}{m_W^2}, \\
\alpha U &= 4 s_W^2 
\left[
\Pi'_{11}(0) - \Pi'_{33}(0)
\right],
\end{align}
\noindent where the prime denotes differentiation with respect to $q^2$, evaluated at $q^2=0$. Here, $\alpha = \frac{e^2}{4\pi}$ and $e^2 = g^2 s_W^2$. The mixed vacuum polarization function $\Pi_{3Q}(q^2)$ corresponds to the correlator between the third component of the weak isospin current, $J_\mu^3$, and the electromagnetic current, $J_\nu^{Q}$.

\par The scalar contribution to the vacuum polarization function $\Pi_{11}(q^2)$ is obtained from the transverse part of the current--current correlator defined in Eq.~(\ref{pola}), and evaluated through the one-loop Feynman diagram showed in Fig.~1. Within this setup, the scalar contributions of the minimal $331\text{RHN}$ model to the oblique parameters can be summarized as
\begin{align}
\alpha S  &=
-\frac{e^2}{48\pi^2}\log\!\left[\frac{m^2_{H^+_{m}}}{m^2_{H_m}}\right] \label{parS} \\
\alpha T  &= \frac{e^2}{64\pi^2 s_W^2 m_W^2}
F(m^2_{H^+_{m}},m^2_{H_m})  \label{parT} \\
\alpha U  &= \frac{e^2}{48\pi^2}
G(m^2_{H^+_{m}},m^2_{H_m}) ,
\label{parU}
\end{align}
\noindent where the loop functions $F(m_1^2,m_2^2)$ and $G(m_1^2,m_2^2)$ are defined as
\begin{align}
F(m_1^2,m_2^2) &=
\frac{m_1^2 + m_2^2}{2}
-
\frac{m_1^2 m_2^2}{m_1^2 - m_2^2}
\ln\!\left(\frac{m_1^2}{m_2^2}\right), \nonumber \\
G(m_1^2,m_2^2) &=
-\frac{5}{6}
+ \frac{m_1^2 m_2^2}{(m_1^2 - m_2^2)^2}
+ \frac{m_1^4 (m_1^2 - 3 m_2^2)}{(m_1^2 - m_2^2)^3}
\ln\!\left(\frac{m_1^2}{m_2^2}\right).
\label{funcFeG}
\end{align}

\subsection{Numerical Results}

\par Non-vanishing contributions to the oblique parameters arise from sectors extending beyond the Standard Model. Within the framework of the minimal $331\text{RHN}$ model, as discussed in Sec.~II-B, the additional quarks $(U_1, D_i)$ introduced in the triplet representations of Eq.~(\ref{quarks}) are singlets under $SU(2)_L$ and therefore do not induce direct contributions to the electroweak oblique parameters.

\par Regarding the gauge sector, the effects associated with the extra gauge bosons $(Y^\pm, X^0, X^{0*}, Z')$ are strongly suppressed and remain well below the current experimental sensitivity to the oblique parameters $S$, $T$, and $U$. This behavior is consistent with global fits to electroweak precision data, which indicate that these observables are tightly constrained around their Standard Model predictions.

\par As emphasized in the Introduction, our previous analysis of the 331RHN model~\cite{331STU:2026dc} demonstrated that the oblique parameter (T) provides the most stringent electroweak constraint on both the scalar mass spectrum and the symmetry-breaking scales of the model. According to the latest global electroweak fit~\cite{PDG}, the parameter (T) is constrained to

\begin{align}
T = 0.01 \pm 0.12,
\label{Tlim}
\end{align}

which we adopt as the primary experimental input throughout our analysis.

\par To identify the regions of parameter space compatible with this constraint, we perform numerical scans over the multidimensional parameter set $(u, \omega, \lambda_1, \lambda_2, \lambda_3, \lambda_4)$, as defined by the scalar mass relations given in Eqs.~(\ref{mhH}) and (\ref{m+1}).
\par In order to reproduce the observed Higgs boson properties, the mass of the Standard Model-like Higgs  is fixed to $(m_h=125.3\mathrm{GeV})$ , as prescribed by Eq.~(\ref{mhH}). 
In our numerical analysis, the parameters are randomly sampled within the ranges

\begin{align}
& 0.001 \leq u \leq 5~\text{GeV}, \nonumber \\
& 500~\text{GeV} \leq \omega \leq 20~\text{TeV}, \nonumber \\
& 0 < \lambda_1, \lambda_2, \lambda_3, \lambda_4 \leq 5.
\end{align}


\par The results of the scan are presented in Fig.~2. The upper panels display the values of the neutral scalar mass $m_{H_m}$, while the lower panels correspond to the charged scalar mass $m_{H^\pm_{m}}$ in the $(u, \omega)$ plane.

\begin{figure}[t]
    \centering
\includegraphics[scale=0.4]{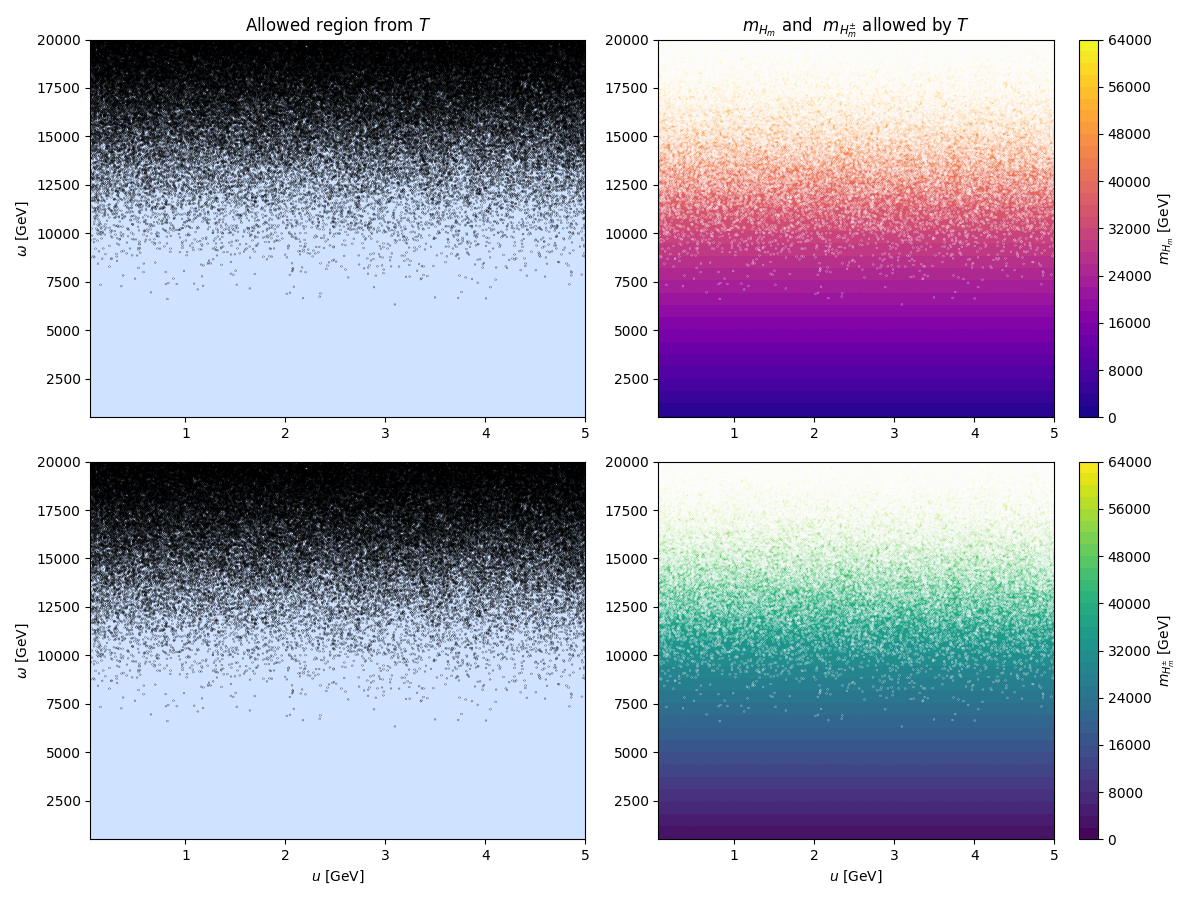}
    \caption{
Predictions of the minimal $331\text{RHN}$ model for the oblique parameter $T$ in the multidimensional parameter space $(u, \omega, \lambda_i)$. 
The left panel shows the regions compatible with the experimental bounds on $T$, while the right panel displays the corresponding scalar mass spectrum, including $m_{H_m}$ and $m_{H^\pm_{m}}$. The white contour indicates the boundary where the oblique parameter reaches its experimental upper limit.}
\end{figure}

\par In each panel, the light-blue region corresponds to the parameter space consistent with the experimental constraint on $T$, while the white contour delineates the boundary at which the oblique parameter saturates its upper limit. The region below this boundary therefore defines the phenomenologically viable domain.

\par The color bar encodes the magnitude of the scalar masses across the scanned parameter space. Although the scan allows values as large as $\omega \simeq 20~\text{TeV}$, the imposition of the $T$ constraint effectively restricts the viable region to $\omega \lesssim 10~\text{TeV}$. Within this domain, the masses of the heavy scalar states remain bounded by the symmetry-breaking scale, $m \lesssim \omega$, in agreement with their parametric dependence on $\omega$.

\par Similarly to what was observed in the  $331\text{RHN}$ model~\cite{331STU:2026dc}
, we find that the oblique parameter $T$ constitutes the most restrictive electroweak constraint on the minimal $331\text{RHN}$ framework, leading to an upper bound of $\omega \lesssim 10~\text{TeV}$ on the symmetry-breaking scale. For completeness, we have also evaluated the scalar contributions to the oblique parameters $S$ and $U$, finding that they remain negligible throughout the allowed parameter space.

\section{Conclusion}

\par In this work, we have performed a systematic analysis of the scalar-sector contributions of the minimal 331RHN model to the oblique parameters $S$, $T$, and $U$. In close analogy with the results previously obtained in  331RHN framework~\cite{331STU:2026dc}, we find that the scalar sector of the minimal realization is likewise subject to stringent constraints from electroweak precision observables. In particular, our results show that the scalar sector of 331RHN minimal model too is strongly constrained by the oblique  parameter $T$ and 
leads to a  upper bound on the symmetry-breaking scale, namely \(\omega \lesssim 10~\text{TeV}
\).

\par In conclusion, the oblique parameter $T$ emerges as the most restrictive electroweak observable within the 331 framework,  providing the strongest bounds on the viable parameter space of these models. These findings reinforce the crucial role of electroweak precision tests as powerful probes of physics beyond the Standard Model.

 \section*{Acknowledgments}
I would like to thank Carlos A. de S. Pires for reading the manuscript and for useful suggestions.


\section*{References}
\bibliography{references}


\end{document}